\documentclass[twocolumn,aps,prd,preprintnumbers,nofootinbib]{revtex4-2}
\usepackage{graphicx}
\usepackage{amsmath}
\usepackage{amssymb}
\usepackage{amsfonts}
\usepackage{hyperref}
\usepackage{url}
\usepackage{color}
\usepackage{bm}

\newcommand{\be}{\begin{equation}}
\newcommand{\ee}{\end{equation}}
\newcommand{\ba}{\begin{eqnarray}}
\newcommand{\ea}{\end{eqnarray}}
\newcommand{\nn}{\nonumber}

\renewcommand{\[}{\begin{equation}}
\renewcommand{\]}{\end{equation}}

\def\lcdm{$\Lambda$CDM }
\def\snt#1{\textcolor{black}{#1}}

\usepackage{array}
\makeatother

\begin{document}

\preprint{IFT-UAM/CSIC-2021-136}

\title{A GREAT model comparison against the cosmological constant}

\author{Rub\'{e}n Arjona}
\email{ruben.arjona@uam.es}

\author{Llorenc Espinosa-Portales}
\email{llorenc.espinosa@uam.es}

\author{Juan Garc\'{i}a-Bellido}
\email{juan.garciabellido@uam.es}

\author{Savvas Nesseris}
\email{savvas.nesseris@csic.es}

\affiliation{Instituto de F\'isica Te\'orica UAM-CSIC, Universidad Auton\'oma de Madrid,
Cantoblanco, 28049 Madrid, Spain}

\date{\today}

\begin{abstract}
Recently, a covariant formulation of non-equilibrium phenomena in the context of General Relativity was proposed in order to explain from first principles the observed accelerated expansion of the Universe, without the need for a cosmological constant, leading to the GREA theory. Here, we confront the GREA theory against the latest cosmological data, including type Ia supernovae, baryon acoustic oscillations, the cosmic microwave background (CMB) radiation, Hubble rate data from the cosmic chronometers and the recent $H_0$ measurements. We perform Markov Chain Monte Carlo analyses and a Bayesian model comparison, by estimating the evidence via thermodynamic integration, and find that when all the aforementioned data are included, but no prior on $H_0$, the difference in the log-evidence is $\sim -9$ in favor of GREA, thus resulting in overwhelming support for the latter over the cosmological constant and cold dark matter model ($\Lambda$CDM). When we also include priors on $H_0$, either from Cepheids or the Tip of the Red Giant Branch measurements, then due to the tensions with CMB data the GREA theory is found to be statistically equivalent with $\Lambda$CDM.
\end{abstract}
\maketitle

\section{Introduction}

Our understanding of the expanding universe is anchored in the geometric description provided by Einstein’s theory of General Relativity (GR). On the one hand, its approximate symmetries, i.e., homogeneity and isotropy at large scales, determine its background space-time to be described by a Friedmann-Lemaître-Robertson-Walker (FLRW) metric. On the other hand, its matter content is responsible for the dynamics of the scale factor, which tracks the growth of length-scales in the geometric expansion, as described by the Friedmann equations.

The currently accepted realization of FLRW cosmology is given by the $\Lambda$ – Cold Dark Matter ($\Lambda$CDM) model. According to it, baryonic matter and radiation make up only a small portion of the present content of the universe. Instead, its expansion is dominated by two components which lack a fully satisfactory microscopic description. First, a cosmological constant, usually denoted by $\Lambda$, which is added to Einstein’s field equations to account for the observed late-time accelerated expansion of the universe. Second, cold (low temperature) dark (without electromagnetic interactions) matter, which was required originally to explain anomalies in the galactic rotation curves but is nowadays consistent with many other early- and late-time cosmological observables.

Even though $\Lambda$CDM seems to be the best fit to observations, the existence of a cosmological constant has been challenged on theoretical grounds. Consequently, a plethora of alternatives haven been explored, which fall systematically into two groups. First, modified gravity (MG) theories attempt to deliver new dynamics at large, cosmological, scales, while leaving invariant smaller scales at which GR has been thoroughly probed. Second, dark energy (DE) models propose the addition of exotic matter, such as quintessence.

Furthermore, in the last years there have been observational challenges to $\Lambda$CDM. Early- and late-time measurements of the present-value of the Hubble parameter ($H_0$) seem to be inconsistent \cite{Riess:2019cxk}. This $H_0$ tension signals a possible failure of the $\Lambda$CDM to describe our universe. However, no available alternative MG or DE seems to be able to resolve the tension between high and low redshift probes, while providing a fit to cosmological observations that is competitive with $\Lambda$CDM \cite{Planck:2018vyg,Kunz:2015oqa}. Moreover, there have been recent model-independent analyses, using machine learning approaches, that suggest that there maybe hints of deviations from \lcdm at high redshifts \cite{Arjona:2020kco,Arjona:2019fwb}.

Recently, a first-principles explanation of cosmic acceleration has been proposed by two of us. This is the General Relativistic Entropic Acceleration (GREA) theory \cite{Garcia-Bellido:2021idr}. It is not based on MG or DE. Rather, it is based on the covariant formulation of non-equilibrium formulation of thermodynamics \cite{Espinosa-Portales:2021cac}. Entropy production during irreversible processes necessarily has an impact on Einstein field equations. This suggests the idea that entropy production or, equivalently, information coarse graining, gravitates. As such, it affects the space-time geometry.

In FLRW cosmology, irreversible processes inevitably contribute with an acceleration term to the Friedmann equations. In GREA, it is the sustained growth of the entropy associated with the cosmic horizon in open inflation scenarios that explains current cosmic acceleration.

The goal of this paper is to test the full viability of the GREA theory at the background level and compare it with the $\Lambda$CDM, against available cosmological data. To that end we consider several datasets: type Ia supernovae, baryon acoustic oscillations (BAO), cosmic  microwave  background  radiation (CMB) and recent determinations of $H_0$. We find that, when all of them are included and no prior on $H_0$ is assumed, Bayesian evidence strongly favors the GREA theory, with difference in log-evidence $\sim 9$. It is to our knowledge the first time an alternative to $\Lambda$CDM performs so ramarkably.  When priors on $H_0$ are included, however, GREA is statistically equivalent to a cosmological constant and future precision tests are required.

This paper is organized as follows. In Sec.~\ref{sec:GREA} we review the the covariant formulation of non-equilibrium thermodynamics and the GREA theory built thereupon. In Sec.~\ref{sec:data} we describe the cosmological data used in our analysis. In Secs.~\ref{sec:MCMC} and \ref{sec:results} we present our results. We finish with our conclusions in Sec.~\ref{sec:conclusions}.

\section{The GREA theory}
\label{sec:GREA}
\subsection{Entropic forces in General Relativity}

The GREA theory \cite{Garcia-Bellido:2021idr} is build upon the covariant formulation of non-equilibrium thermodynamics in GR \cite{Espinosa-Portales:2021cac}. This formalism provides a rigorous synthesis of the variational formulation of GR and the second law of thermodynamics. As a result, it predicts the emergence of entropic forces associated to any out-of-equilibrium phenomenon, i.e., any increase in entropy. The Einstein field equations are modified by the introduction of term that encodes such a force
\begin{equation}
    R_{\mu \nu} - \frac{1}{2} R g_{\mu \nu} = 8\pi G \left( T_{\mu \nu} - f_{\mu \nu}\right)\,,
\end{equation}
where $f_{\mu \nu}$ is the entropic force tensor. Its precise form is obtained in the Arnowitt-Deser-Misneer (ADM) formalism from the relation between the time evolution of the spatial metric and the local production of entropy. When applied to homogeneous and isotropic cosmology, it leads to the modified Friedmann equations
\begin{equation}
\begin{aligned}
    H^2 & = \frac{8\pi G}{3} \rho - \frac{k}{a^2}, \\
     \frac{\ddot{a}}{a} & = - \frac{4\pi G}{3} \left( \rho + 3p - \frac{T\dot{S}}{H a^3}\right)\,.
\end{aligned}
\end{equation}
In this setup, the cosmic fluid satisfies the out-of-equilibrium continuity equation
\begin{equation}
    \dot{\rho} + 3 H (\rho + p) = \frac{T \dot{S}}{a^3}\,.\label{eq:conti}
\end{equation}
One concludes from the form of the entropic force in the second Friedmann equation that entropy production leads in general to a \emph{positive} contribution to the acceleration of the universe.

There are two sources of entropy that fit naturally in the variational formalism. On the one hand, the matter Lagrangian may depend on the entropy or entropy density. We call this \textit{bulk entropy}. On the other hand, one may be assign entropy to horizons, as inspired by black hole thermodynamics. This is achieved by adding a Gibbons-Hawking-York (GHY) term that is then interpreted as thermodynamic contribution to the action. We call this \textit{boundary entropy}.

\textit{Bulk entropy} is produced during cosmic expansion during certain out-of-equilibrium processes, such as (p)reheating, phase transitions or gravitational collapse. However, most of the expansion history of the universe is adiabatic and deviations from it are expected to be short-lived. This means that, although it may provide interesting phenomenology, it seems unable to explain the current accelerated expansion of the universe. On the contrary, \textit{boundary entropy} can undergo a sustained increase that becomes relevant only at recent times.

\subsection{Cosmic acceleration from boundary entropy}

Let  us  consider  an  open  universe nucleated in  de Sitter  space,  i.e.  in  eternal  inflation~\cite{Linde:1999wv}.  Inside the true vacuum bubble, local  space-time  as  seen  by  a  comoving  observer  is  essentially  flat  if  inflation  lasts  long enough, e.g.  of order $N\sim 70$ e-folds.  Nevertheless, the bubble walls are still located at a finite coordinate distance and, thus, we can define a true casual horizon with $\sqrt{-k} = a_0H_0$.  Inspired by this scenario we propose a GHY thermodynamic term that induces an entropic contribution satisfying~\cite{Garcia-Bellido:2021idr} 
\ba\label{eq:Causal}
\rho_H\,a^2 = \frac{T_H S_H}{a} &=& \frac{1}{2G}\,
\frac{\sinh(2a_0H_0\eta)}{a_0H_0}\,,\\[2mm]
\frac{\Omega_K}{1-\Omega_K} &=& e^{-2N}
\left(\frac{T_{\rm rh}}{T_{\rm eq}}\right)^2(1+z_{\rm eq})\,,
\ea
where $\eta$ is the conformal time, $\Omega_K$ is the curvature parameter inside the inflated patch, $T_{\rm rh}$ is the reheating temperature, $T_{\rm eq}$ and $z_{\rm eq}$ are, respectively, the temperature and redshift at matter-radiation equality.  We now introduce, for convenience, the time coordinate $\tau = a_0 H_0 \eta$ and denote with primes the derivatives w.r.t. to $\tau$. Then the second Friedmann equation becomes
\begin{equation}
\left(\frac{a'}{a_0}\right)^2 \!= \Omega_M\,\frac{a}{a_0} + \Omega_K\,\frac{a^2}{a_0^2} +  \frac{4\pi}{3}\,\Omega_K^{3/2}\,\frac{a^2}{a_0^2}\,\sinh(2\tau) \,,\label{eq:friedGreat}
\end{equation}
where $\Omega_M$ is the matter density parameter.

Thus, the expansion of the universe is affected by the increase in entropy of the causal horizon. Since the causal horizon keeps growing, the entropic term eventually dominates and leads to a late-time cosmic acceleration. Contrary to a cosmological constant, however, the entropic term is diluted with the expansion, albeit at a slower rate than radiation and dust, and the universe ends in Minkowski space-time in the far future.

From the mathematical point of view, this modified second Friedmann equation is a differential equation in re-scaled conformal time $\tau$. It is, however, an integro-differential equation in cosmic time $t$, unlike the usual second Friedmann equation. Physically, this is related to the nature of the entropic term associated to the causal horizon: it builds up as the expansion proceeds.

\section{The data}
\label{sec:data}
Here we present in detail the compilations of data we use in our analysis.
\subsection{The $H(z)$ data \label{sec:hzdata}}
First, we consider the Hubble rate data, which are obtained via two complementary ways. The first one is from the redshift drift of distant objects over long periods of time, usually on the order of a decade. This is possible as in the FRLW metric the Hubble parameter $H(z)$ can be related to the rate of change of redshift with respect to time, i.e. $H(z)=-\frac{1}{1+z}\frac{dz}{dt}$ \cite{Jimenez:2001gg}. In particular, the $H(z)$ data are determined via the differential age method using the evolution of $D_n4000$, which is a spectral feature of very massive and passive galaxies. The systematics in this case mainly come from the metallicity, via the M11 and BC03 models discussed in Ref.~\cite{Moresco:2016mzx}. However, it has been shown that the systematics can be kept under control by implementing strict selection criteria \cite{Moresco:2016mzx}. 

On the other hand, some measurements also come from the  clustering of galaxies or quasars, which is a probe of the Hubble expansion via the determination of the BAO in the radial direction \cite{Gaztanaga:2008xz}. Furthermore, we assume that the $H(z)$ data are uncorrelated with each other. Finally, here we will make use of the compilation from Ref.~\cite{Arjona:2018jhh} that contains $36$ points in the redshift range $0.07\le z \le 2.34$ and which are in the form $(z_i,H_i,\sigma_{H_i})$, as is shown in Table~\ref{tab:Hzdata}. 

\begin{table}[!t]
\caption{The $H(z)$ data used in our analysis (in units of $\textrm{km}~\textrm{s}^{-1} \textrm{Mpc}^{-1}$). This compilation, which was presented in Ref.~\cite{Arjona:2018jhh}, is partly based on those of Refs.~\cite{Moresco:2016mzx} and \cite{Guo:2015gpa}.\label{tab:Hzdata}}
\centering
\begin{tabular}{cccccccccc}
\\
\hline\hline
$z$  & $H(z)$ & $\sigma_{H}$ & Ref.   \\
\hline
$0.07$    & $69.0$   & $19.6$  & \cite{Zhang:2012mp}  \\
$0.09$    & $69.0$   & $12.0$  & \cite{STERN:2009EP} \\
$0.12$    & $68.6$   & $26.2$  & \cite{Zhang:2012mp}  \\
$0.17$    & $83.0$   & $8.0$   & \cite{STERN:2009EP}    \\
$0.179$   & $75.0$   & $4.0$   & \cite{MORESCO:2012JH}   \\
$0.199$   & $75.0$   & $5.0$   & \cite{MORESCO:2012JH}   \\
$0.2$     & $72.9$   & $29.6$  & \cite{Zhang:2012mp}   \\
$0.27$    & $77.0$   & $14.0$  & \cite{STERN:2009EP}   \\
$0.28$    & $88.8$   & $36.6$  & \cite{Zhang:2012mp}  \\
$0.35$    & $82.7$   & $8.4$   & \cite{Chuang:2012qt}   \\
$0.352$   & $83.0$   & $14.0$  & \cite{MORESCO:2012JH}   \\
$0.3802$  & $83.0$   & $13.5$  & \cite{Moresco:2016mzx}   \\
$0.4$     & $95.0$   & $17.0$  & \cite{STERN:2009EP}    \\
$0.4004$  & $77.0$   & $10.2$  & \cite{Moresco:2016mzx}   \\
$0.4247$  & $87.1$   & $11.2$  & \cite{Moresco:2016mzx}   \\
$0.44$    & $82.6$   & $7.8$   & \cite{Blake:2012pj}   \\
$0.44497$ & $92.8$   & $12.9$  & \cite{Moresco:2016mzx}   \\
$0.4783$  & $80.9$   & $9.0$   & \cite{Moresco:2016mzx}   \\
\hline\hline
\end{tabular}~~~~~~~~
\begin{tabular}{cccccccccc}
\\
\hline\hline
$z$  & $H(z)$ & $\sigma_{H}$ & Ref.   \\
\hline
$0.48$    & $97.0$   & $62.0$  & \cite{STERN:2009EP}   \\
$0.57$    & $96.8$   & $3.4$   & \cite{Anderson:2013zyy}   \\
$0.593$   & $104.0$  & $13.0$  & \cite{MORESCO:2012JH}  \\
$0.60$    & $87.9$   & $6.1$   & \cite{Blake:2012pj}   \\
$0.68$    & $92.0$   & $8.0$   & \cite{MORESCO:2012JH}    \\
$0.73$    & $97.3$   & $7.0$   & \cite{Blake:2012pj}   \\
$0.781$   & $105.0$  & $12.0$  & \cite{MORESCO:2012JH} \\
$0.875$   & $125.0$  & $17.0$  & \cite{MORESCO:2012JH} \\
$0.88$    & $90.0$   & $40.0$  & \cite{STERN:2009EP}   \\
$0.9$     & $117.0$  & $23.0$  & \cite{STERN:2009EP}   \\
$1.037$   & $154.0$  & $20.0$  & \cite{MORESCO:2012JH} \\
$1.3$     & $168.0$  & $17.0$  & \cite{STERN:2009EP}   \\
$1.363$   & $160.0$  & $33.6$  & \cite{Moresco:2015cya}  \\
$1.43$    & $177.0$  & $18.0$  & \cite{STERN:2009EP}   \\
$1.53$    & $140.0$  & $14.0$  & \cite{STERN:2009EP}  \\
$1.75$    & $202.0$  & $40.0$  & \cite{STERN:2009EP}  \\
$1.965$   & $186.5$  & $50.4$  & \cite{Moresco:2015cya}  \\
$2.34$    & $222.0$  & $7.0$   & \cite{Delubac:2014aqe}   \\
\hline\hline
\end{tabular}
\end{table}

\subsection{The SnIa data}
We also use the Pantheon supernovae type Ia data (SnIa) compilation of Ref.~\cite{Scolnic:2017caz} of 1048 Supernovae Ia points in the redshift range $0.01<z<2.26$, along with their covariance matrix. The apparent magnitude $m_B$ of the SnIa points is given by
\be
m_\textrm{B}=5 \log_{10}\left[\frac{D_\textrm{L}(z)}{1 \textrm{Mpc}}\right]+25+M_B,
\ee
where $D_\textrm{L}(z)$ is the luminosity distance and $M_B$ the absolute magnitude. Finally, the parameter $M_B$ is marginalized over, according to the recipe in Appendix C of Ref.~\cite{Conley:2011ku}. 

\subsection{The BAO}
The compilation of BAO data used in our analysis includes points from 6dFGS \cite{Beutler:2011hx}, WiggleZ \cite{Blake:2012pj}, the MGS, ELG, LRG, quasars and DR12 galaxy samples BAO points from the completed SDSS-IV eBOSS survey \cite{eBOSS:2020yzd}, the year 3 DES \cite{DES:2021esc} and the Lyman-$\alpha$ (Ly$\alpha$) absorption and quasars, auto and cross correlation points from  Ref.~\cite{duMasdesBourboux:2020pck}.

In what follows, we will briefly discuss the functions which are used to describe the BAO data. A key quantity is the ratio of the sound horizon at the drag redshift $r_s(z_d)$ to the so-called dilation scale $D_V(z)$: \be
d_z\equiv \frac{r_s(z_d)}{D_V(z)},\label{eq:dz}
\ee
where the comoving sound horizon is 
\be
r_s(z_d)=\int_{z_d}^{\infty}\frac{c_s(z)}{H(z)}dz,\label{eq:rsd}
\ee 
where the redshift at the dragging epoch $z_d$ is given for example by Eq.~(4) of \cite{Eisenstein:1997ik}, however to actually evaluate the integral of Eq.~\eqref{eq:rsd} we will use the fitting formula from Ref.~\cite{Aizpuru:2021vhd}, which is obtained via machine learning improved fits of the full recombination history, resulting in 
\be
z_\mathrm{d}=\frac{1+428.169 \omega_b^{0.256459}\omega_m^{0.616388}+925.56 \omega_m^{0.751615}}{\omega_m^{0.714129}},
\ee
and which is accurate up to $\sim 0.001\%$ \cite{Aizpuru:2021vhd}.
In Eq.~\eqref{eq:dz} we also defined the dilation scale $D_V(z)$, which  is given by
\be
D_V(z)=\left[(1+z)^2 D_A(z)^2 \frac{c z}{H(z)}\right]^{1/3},
\ee
where $D_A(z)$ is the angular diameter distance. Finally, we can also define the Hubble and comoving angular diameter distances, via 
\ba
D_H(z)&=&c/H(z),\\
D_M(z)&=&(1+z) D_A(z).
\ea

Next we describe the actual BAO data. In particular, the 6dFGs and WiggleZ points are given by
\be
\begin{array}{ccc}
 z  & d_z & \sigma_{d_z } \\
 \hline
 0.106 & 0.336 & 0.015 \\
 0.44 & 0.073 & 0.031 \\
 0.60 & 0.0726 & 0.0164 \\
 0.73 & 0.0592 & 0.0185 \\
\end{array}
\ee
where their inverse covariance matrix is
\be C_{ij}^{-1}=\left(
\begin{array}{cccc}
 \frac{1}{0.015^2} & 0 & 0 & 0 \\
 0 & 1040.3 & -807.5 & 336.8 \\
 0 & -807.5 & 3720.3 & -1551.9 \\
 0 & 336.8 & -1551.9 & 2914.9 \\
\end{array}
\right)\ee
with the $\chi^2$ is then given by
\be
\chi^2_\textrm{6dFS,Wig}=V^i\;C_{ij}^{-1}\; V^j,\label{eq:chi26df}
\ee
where the difference vector is given by $V^i=d_{z,i}-d_z(z_i)$. 

The BAO measurements for MGS and eBOSS ELGs are given by $D_V/r_s = 1/d_z$ via
\be
\begin{array}{ccc}
 z  & 1/d_z & \sigma_{1/d_z } \\
 \hline
 0.15 & 4.46567 & 0.168135 \\
 0.85 & 18.33 & 0.595 \\
\end{array}
\ee
and the $\chi^2$ is
\be
\chi^2_\textrm{MGS,ELG}=\sum_{i=1}^2 \left[\frac{1/d_{z,i}-1/d_z(z_i)}{\sigma_{1/d_{z,i}}}\right]^2.\label{eq:chi2SDSS}
\ee

The BAO data from DES year 3 are of the form $D_M(z)/r_s$ with $\left[z,D_M(z)/r_s,\sigma_{D_{M,i}/r_s}\right]=(0.835, 18.92, 0.51)$ and the $\chi^2$ given by \be
\chi^2_\textrm{DES}= \left[\frac{D_{M,i}/r_s-D_M(z_i)/r_s}{\sigma_{D_{M,i}/r_s}}\right]^2.
\ee

We also include the eBOSS LRG data, which are given by $\left(z,D_M/r_s,D_H/r_s\right)=(0.698,17.8581,19.3261)$ with an inverse covariance matrix
\be 
C_{ij}^{-1}=\left(
\begin{array}{cc}
 10.4515 & 2.14754 \\
 2.14754 & 3.96466 \\
\end{array}
\right),
\ee
so that the $\chi^2$ is 
\be
\chi^2_\mathrm{LRG}=V^i\, C_{ij}^{-1}\,V^j,
\ee 
where the difference vector is 
\be 
V^i=\left[D_{M,i}-D_{M}(z_i),D_{H,i}-D_{H}(z_i)\right]/r_s.
\ee

Similarly the eBOSS QSO points are given by 
$\left(z,D_M/r_s,D_H/r_s\right)=(1.48,30.6876,13.2609)$ with an inverse covariance matrix
\be 
C_{ij}^{-1}=\left(
\begin{array}{cc}
 1.84606 & -1.0342 \\
 -1.0342 & 3.86146 \\
\end{array}
\right),
\ee
so that the $\chi^2$ is 
\be
\chi^2_\mathrm{QSO}=V^i\, C_{ij}^{-1}\,V^j,
\ee 
where the difference vector is \be 
V^i=\left[D_{M,i}-D_{M}(z_i),D_{H,i}-D_{H}(z_i)\right]/r_s.
\ee

We also include the BAO data from Ly$\alpha$ and the cross/auto correlations with the quasars, which are of the form $f_\textrm{BAO}=(D_H/r_s,D_M/r_s)$ and are given by
\be
\begin{array}{ccc}
 z & f_\textrm{BAO}  &  \sigma_{f_\textrm{BAO}} \\
 \hline
 2.334 & 8.99 & 0.429418 \\
 2.334 & 37.5 & 2.77308 \\
\end{array}
\ee
with a correlation coefficient $\rho=-0.45$, so that the $\chi^2$ given by 
\be
\chi^2_\mathrm{Lya}=V^i\, C_{ij}^{-1}\,V^j,
\ee 
where the difference vector is \be 
V^i=\left[D_{M,i}-D_{M}(z_i),D_{H,i}-D_{H}(z_i)\right]/r_s.
\ee

Finally, the eBOSS DR12 galaxy samples data are of the form $f_\textrm{BAO}=(D_M/r_s,D_H/r_s)$ and are given by 
\be
\begin{array}{ccc}
 z  & D_M/r_s & D_H/r_s \\
 \hline
 0.38 & 10.2341 & 24.9806 \\
 0.51 & 13.366 & 22.3166 \\
\end{array}
\ee
with an inverse covariance matrix
\be 
C_{ij}^{-1}=
\left(
\begin{array}{cccc}
 52.584 & 5.15947 & -20.0391 &
   -3.54599 \\
 5.15947 & 2.8048 & -2.10831 &
   -1.61178 \\
 -20.0391 & -2.10831 & 36.8787 &
   5.7886 \\
 -3.54599 & -1.61178 & 5.7886 &
   4.64349 \\
\end{array}
\right),
\ee 
while the difference vector is 
\ba
V^i&=&\left[D_{M,0.38},D_{H,0.38},D_{M,0.51},D_{H,0.51}\right]/r_s\nn \\
&-&\left[D_{M}(0.38),D_{H}(0.38),D_{M}(0.51),D_{H}(0.51)\right]/r_s,\nn \\
\ea 
with the $\chi^2$ given by
\be
\chi^2_\mathrm{DR12}=V^i\, C_{ij}^{-1}\,V^j.
\ee 

Finally, the total $\chi^2$ is then given by
\ba
\chi^2_\textrm{BAO}&=&\chi^2_\textrm{6dFS,Wig}+ \chi^2_\textrm{MGS,ELG} + \chi^2_\textrm{DES}+ \chi^2_\textrm{LRG}+ \chi^2_\textrm{QSO}\nn \\
&+&\chi^2_\textrm{Lya}+\chi^2_\textrm{DR12}.\label{eq:chi2tot}
\ea
Note that in the latter equation we assume that the data are independent with each other, thus we can simply add the $\chi^2$ terms. However, since some of the points are derived by the same survey, inevitably there will be common overlapping galaxies between the datasets, which will result to strong covariances, which is clearly a limitation in our analysis. 

For example for the WiggleZ data the correlations between the points is given by the covariance matrix $C_{ij}$, thus we have included this information in our analysis. However, overall the full correlations are  not publicly available and it is impossible to correctly estimate a covariance matrix, even if a few attempts have been made in the literature, e.g. for a similar discussion for the growth-rate data see Ref.~\cite{Alam:2015rsa}. 

\subsection{The CMB shift parameters}
\snt{The main effects of the new entropy terms will be twofold: one on the background Friedmann equation given by Eq.~\eqref{eq:friedGreat} and another on possible contributions to the perturbations as seen by Eq.~\eqref{eq:conti}. Currently, a perturbation theory for the GREA model is not readily available, thus in this work we only focus on the background contributions and leave the full perturbation analysis for future work.
}

Thus, we can use the so called CMB shift parameters \cite{Wang:2007mza,Zhai:2018vmm}. Furthermore, this simplifies the analysis as most Boltzmann codes calculate the conformal time, after having calculated the Hubble parameter, which make modifications of codes like \texttt{CAMB} or \texttt{CLASS} highly non-trivial. The CMB shift parameters encapsulate the geometric information in the CMB spectrum, via the location of the peaks and are in a sense a compressed form of the CMB likelihood. They are given by
\ba 
R &\equiv& \sqrt{\Omega_{m,0}H_0^2}\, r(z_\mathrm{rec})/c, \label{eq:R} \\
l_a&\equiv & \pi \, r(z_\mathrm{rec})/r_s(z_\mathrm{rec}),\label{eq:la} 
\ea 
where $r_s(z_\mathrm{rec})$ is the sound horizon at recombination and $z_\mathrm{rec}$ is the redshift at recombination, which can be calculated by the fitting formula of Ref.~\cite{Aizpuru:2021vhd}. 

As here we are interested in non-flat universes we use the Planck 2018 chains base\_omegak\_plikHM\_TTTEEE\_lowl\_lowE\_lensing to estimate the data vectors for $(R, l_a, \Omega_b h^2, h)$. \snt{Note that the curvature is in fact included in our compressed likelihood as the Planck 2018  chains we used include a free curvature parameter, as denoted by the name of the chain. Thus, the curvature appears directly in the likelihood, since the parameters $R$ and $l_a$ given by Eqs.~\eqref{eq:R}-\eqref{eq:la} depend explicitly on $\Omega_{k}$. Following then the procedure of Refs.~\cite{Wang:2007mza,Zhai:2018vmm} we find } 
\be 
\mathbf{v}=
\left(
  \begin{array}{c}
1.74448\\ 
302.21792\\ 
0.02249\\
0.63549
  \end{array}
\right),
\ee 
while the covariance matrix is 
\ba 
&&\mathbf{C}_{v}= 
10^{-8} \times \nn \\
&&\left(
\begin{array}{cccc}
2604.44383 & 16594.36494 & -58.52126 & 4633.20089 \\
16594.36494 &  738151.92316 & -410.26313 & 20120.28532\\
-58.52126& -410.26313 & 2.58145 & -93.88730\\
4633.20089& 20120.28532 & -93.88730 & 49803.48059\\
\end{array}
\right).\nn \\
\ea 
Thus, the difference vector can be written as 
\be 
\mathbf{V}=[R, l_a, \Omega_b h^2, h]-\mathbf{v},
\ee 
thus, the $\chi^2$ for the CMB data can be written as 
\be 
\chi^2_\mathrm{cmb}=\mathbf{V}\, \mathbf{C}_{v}^{-1}\, \mathbf{V}.
\ee 

\subsection{The Riess $H_0$ prior}
We also use the $H_0$ measurement from Ref.~\cite{Riess:2020fzl}, which comes from a sample of 75 Milky Way Cepheids, which were used to recalibrate the extragalactic distance ladder. This approach gives
\be 
H_0^{(\mathrm{R})}=73.2\pm1.3 \,\mathrm{km}\,\mathrm{s}^{-1}\, \mathrm{Mpc}^{-1}.
\ee 
Then, the $\chi^2$ term is just
\be 
\chi^2_{\rm H_0}=\left(\frac{H_0^{(\mathrm{R})}-H_0}{\sigma_{H_0^{(\mathrm{R})}}}\right)^2,
\ee 
where the Hubble parameter today is given by $H_0=100\,h$ in the \lcdm model and by evaluating Eq.~\eqref{eq:friedGreat} at $\tau=\tau_0$, i.e. at today, for the GREAT model.

\subsection{The TRGB $H_0$ prior}
Finally, we also include the $H_0$ measurement from Ref.~\cite{Freedman:2020dne}, which comes from the Tip of the Red Giant Branch (TRGB) method using stars in the Large Magellanic Cloud (LMC). This approach gives
\be 
H_0^{(\mathrm{TRGB})}=69.6\pm 0.8 \,(\mathrm{stat})\,\pm 1.7 \,(\mathrm{syst})\,\,\mathrm{km}\,\mathrm{s}^{-1}\, \mathrm{Mpc}^{-1}.
\ee 
Then, the $\chi^2$ term is just
\be 
\chi^2_{\rm H_0}=\left(\frac{H_0^{(\mathrm{TRGB})}-H_0}{\sigma_{H_0^{(\mathrm{TRGB})}}}\right)^2,
\ee 
where the Hubble parameter today is given by $H_0=100\,h$ in the \lcdm model and by evaluating Eq.~\eqref{eq:friedGreat} at $\tau=\tau_0$, i.e. at today, for the GREAT model.

\section{MCMC}
\label{sec:MCMC}
In this section we present the results of our Markov Chain Monte Carlo (MCMC) analysis after fitting the data described in Sec.~\ref{sec:data}. Our total likelihood function $L_{\rm tot}$ can be given as the product of the various likelihoods as
$$
\mathcal{L}_{\rm tot}=\mathcal{L}_{\rm SnIa} \times \mathcal{L}_{\rm BAO} \times \mathcal{L}_{\rm H(z)} \times \mathcal{L}_{\rm cmb}\times \mathcal{L}_{\rm H_0},
$$
which can also be translated to the total $\chi^2$ via $\chi^{2}_{\rm tot}=-2\ln{\mathcal{L}_{\rm tot}}$ or
\be
\chi^{2}_{\rm tot}=\chi^{2}_{\rm SnIa}+\chi^{2}_{\rm BAO}+\chi^{2}_{\rm H(z)}+
\chi^{2}_{\rm cmb}+\chi^2_{\rm H_0}.\label{eq:chi2eq}
\ee

Our $\chi^2$ is given by Eq.~(\ref{eq:chi2eq}) and the parameter vectors for both the \lcdm and GREAT models are given by: $p_{ \textrm{Model}}=\left(\Omega_{m0}, \Omega_b h^2, h, \Omega_{k}\right)$. Then, the best-fit parameters and their uncertainties were obtained via an MCMC code written by one of the authors\footnote{\href{https://github.com/snesseris/GREAT-project}{https://github.com/snesseris/GREAT-project}}. Moreover, we assumed priors for the parameters of the \lcdm model given by $\Omega_{m0} \in[0.01, 0.5]$, $\Omega_b h^2 \in[0.015, 0.035]$, $\Omega_{k} \in[-0.1, 0.1]$, $h \in[0.5, 1]$, while for the GREAT model we chose $\Omega_{m0} \in[0.01, 0.5]$, $\Omega_b h^2 \in[0.015, 0.035]$, $\Omega_{k} \in[0.00001, 0.1]$, $h \in[0.5, 1]$\footnote{Note that for the GREAT model $\Omega_{k}$ has to be positive as otherwise the square of the Hubble parameter may become negative. }. Finally, we obtained approximately $\mathcal{O}(10^5)$ points for each of the models.

\begin{table}
\begin{center}
\caption{The values of both the linear and the logarithmic Jeffreys' scale. \label{tab:Jef}} 
\begin{tabular}{cccc}
  \hline
  \hline
  \hspace{5pt} $B_{ij}$ \hspace{5pt} & \hspace{5pt} $\ln{B_{ij}}$ \hspace{5pt}&\hspace{5pt} Evidence \hspace{5pt}\\
  \hline
  \hspace{5pt}$<3$  \hspace{5pt} &\hspace{5pt} $<1.1$ \hspace{5pt}&\hspace{5pt} Weak       \\
  \hspace{5pt}$<20$ \hspace{5pt} &\hspace{5pt} $<3$   \hspace{5pt}&\hspace{5pt} Definite   \\
  \hspace{5pt}$<150$\hspace{5pt} &\hspace{5pt} $<5$   \hspace{5pt}&\hspace{5pt} Strong     \\
  \hspace{5pt}$>150$\hspace{5pt} &\hspace{5pt} $>5$   \hspace{5pt}&\hspace{5pt} Very Strong\\
  \hline
\end{tabular}
\end{center}
\end{table}

In order to compare the quality of fit between the models, we use Bayesian model comparison by means of the evidence $B$. The latter is calculated as the integral of the product of the total likelihood and the priors, over all parameters, that is
\be 
E_i\equiv \int d^nx\,\mathcal{L}_i(x)\, p(x),
\ee 
where $p(x)$ is the prior, while the likelihood for a model $M_i$ is given by $\mathcal{L}_i(x)$ for some parameters $\mathbf{x}$. In practice, as the numerical evaluation of the integral is cumbersome, we use thermodynamic integration following the recipe in Appendix~\ref{sec:app} and Refs.~\cite{Beltran:2005xd,thermoint}. In an nutshell, the temperature rescaled evidence can be written as
\be 
Z(\beta)=\int d^n x\,\mathcal{L}(x)^\beta \, p(x),
\ee 
where $\beta=1/T$ is the inverse temperature and the evidence is given by $E_i\equiv Z_i(1)$, where the latter can be calculated by doing MCMCs at different temperatures and integrating the expectation value of the log-likelihood over the range $\beta\in[0,1]$, see Eq.~\eqref{eq:Z1}.

Then, the comparison of the models is done via the ratio of the evidence for different models, ie
\be
B_{ij}=\frac{E_i}{E_j},
\ee 
which may be interpreted via Jeffreys' scale. The latter can interpret the Bayes ratio as providing evidence in favor of or against  model ${\cal M}_i$ when compared against model ${\cal M}_j$. In a nutshell, every time $\ln B_{ij}$ increases by a unit, this is interpreted as providing further support for one of the two models, with 0 meant as indecisive, to larger than 5 being strongly ruled out. Furthermore, the specific values of the Bayes ratio can be interpreted as follows \cite{John:2002gg}: a value in the range $1<B_{ij}<3$ implies some evidence, which in practice is only barely worth a mention, against $M_j$ when compared with $M_i$. For values in the range $3<B_{ij}<20$ this implies definite but not strong evidence against $M_j$, while for $20<B_{ij}<150$ the evidence is strong and finally, when $B_{ij}>150$ the evidence is very strong. Note however, that it was shown in Ref.~\cite{Nesseris:2012cq} that the Jeffreys' scale has to be interpreted with care, especially in the case of nested models, as it may result to biased conclusions.

Finally, for easy reference we show the particular values of both the linear and the logarithmic Jeffreys' scale in Table \ref{tab:Jef}.

\begin{figure*}[!t]
    \centering
    \includegraphics[width=0.4972\textwidth]{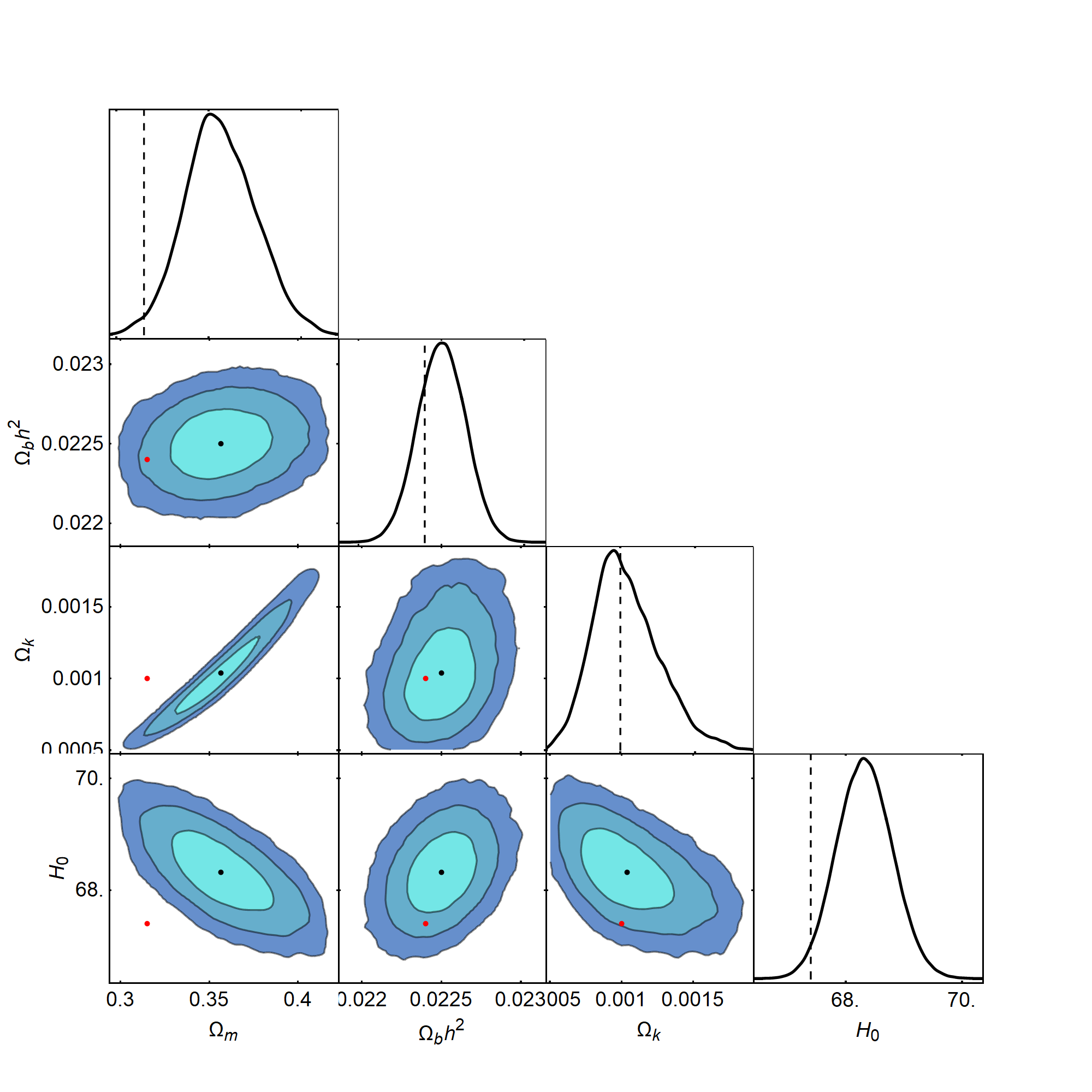}
    \includegraphics[width=0.4972\textwidth]{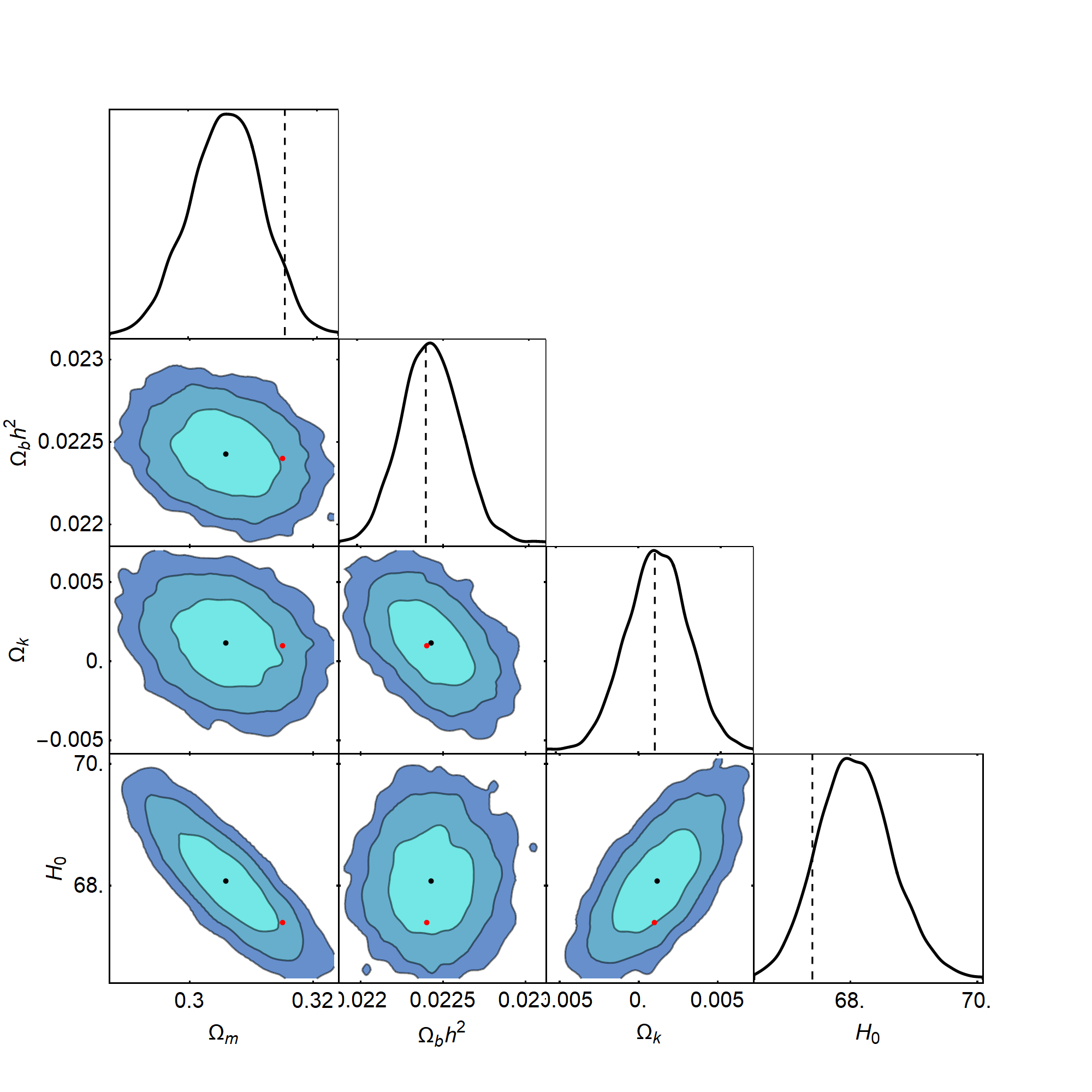}
    \caption{The $68\%$ and $95\%$ confidence contours for the GREAT (left panel) and \lcdm (right panel) models respectively, including all the data, but no prior on $H_0$. The red points/dashed lines correspond to the Planck best-fit $(\Omega_{m,0},\Omega_{b,0}h^2,\Omega_{k,0},H_0)=(0.315,0.0224,0.001,67.4)$, where $H_0$ is given in units of $\mathrm{km}\,\mathrm{s}^{-1}\,\mathrm{Mpc}^{-1}$.}
    \label{fig:contours_no_H0}
\end{figure*}

\begin{table*}[!t]
\begin{tabular}{cccccccc}
  \hline
  Model & $\Omega_{m,0}$ & $\Omega_{b,0}h^2$ & $\Omega_{k,0}$ & $H_0$ & $\chi^2_{min}$ & $\log Z(1)$ &$ \Delta \log Z(1)_{\Lambda,i}$\\ \hline
  $\Lambda$CDM & $0.3057 \pm 0.0056 $ & $0.0224\pm  0.0002$ & $0.0012\pm 0.0018 $ & $68.08\pm0.58 $ & $1075.63 $ & -557.515 & $0$ \\
  GREAT & $0.3522 \pm 0.0190 $ & $0.0225\pm 0.0001 $ & $0.0010\pm 0.0002 $ & $68.38\pm0.48 $ & $1071.35$& -548.509 & $-9.006$ \\
  \hline
\end{tabular}
\caption{Here we present the results of the MCMC analysis when not including any $H_0$ prior. In particular, we show the mean values, $1\sigma$ errors of the parameters for the GREAT and \lcdm models respectively, along with the minimum $\chi^2$ and the log-evidence $\log Z(1)$, see appendix~\ref{sec:app} and the difference of the log-evidence with respect to the \lcdm model $ \Delta \log Z(1)_{\Lambda,i}\equiv\log Z(1)_{\Lambda}-\log Z(1)_{i}$. The latter give a Bayes ratio of $B_{\Lambda, G}=\exp\left[\Delta \log Z(1)_{\Lambda,G}\right]=\exp\left(-9.006\right)\sim 1/8150$, thus resulting in very strong evidence in favor of the GREAT model according to the Jeffreys' scale \cite{Nesseris:2012cq}. Note that $H_0$ is given in units of $\mathrm{km}\,\mathrm{s}^{-1}\,\mathrm{Mpc}^{-1}$. \label{tab:vals_no_H0}  }
\end{table*}

\begin{table*}[!t]
\begin{tabular}{cccccc}
  \hline
  Model 		& CMB	& BAO	& SnIa 	& $H(z)$ & $\chi^2_\mathrm{tot}$ \\ \hline
  $\Lambda$CDM 	& $4.28$ 	& $13.99$ 	& $1034.84$ 	& $22.52$ 	 & $1075.63$\\
  GREAT 		& $0.07$ 	& $14.39$ 	& $1034.82$ 	& $22.10$ 	 & $1071.35$\\
  \hline
\end{tabular}
\caption{The breakdown of the $\chi^2$ for the two models and the different datasets used in our analysis, in the case of not including any $H_0$ prior. The best-fit parameters from the MCMC are given in Table \ref{tab:vals_no_H0}. As can be seen, the main contribution in the difference of the $\chi^2$s comes from the CMB and to a lesser extent from the $H(z)$ and BAO data, while the values for the SnIa are practically the same.\label{tab:vals_no_H0_chi2s}  }
\end{table*}

\section{Results}
\label{sec:results}
Here we present the results of our analysis for both the \lcdm and the GREAT models, using the methodology described in the previous sections. In all cases in the Tables that follow we will show the mean values, $1\sigma$ errors of the parameters for the GREAT and \lcdm models respectively, along with the minimum $\chi^2$, the log-evidence $\log Z(1)$ and the difference of the log-evidence with respect to the \lcdm model $ \Delta \log Z(1)_{\Lambda,i}\equiv\log Z(1)_{\Lambda}-\log Z(1)_{i}$. 

\begin{figure*}[!t]
    \centering
    \includegraphics[width=0.497\textwidth]{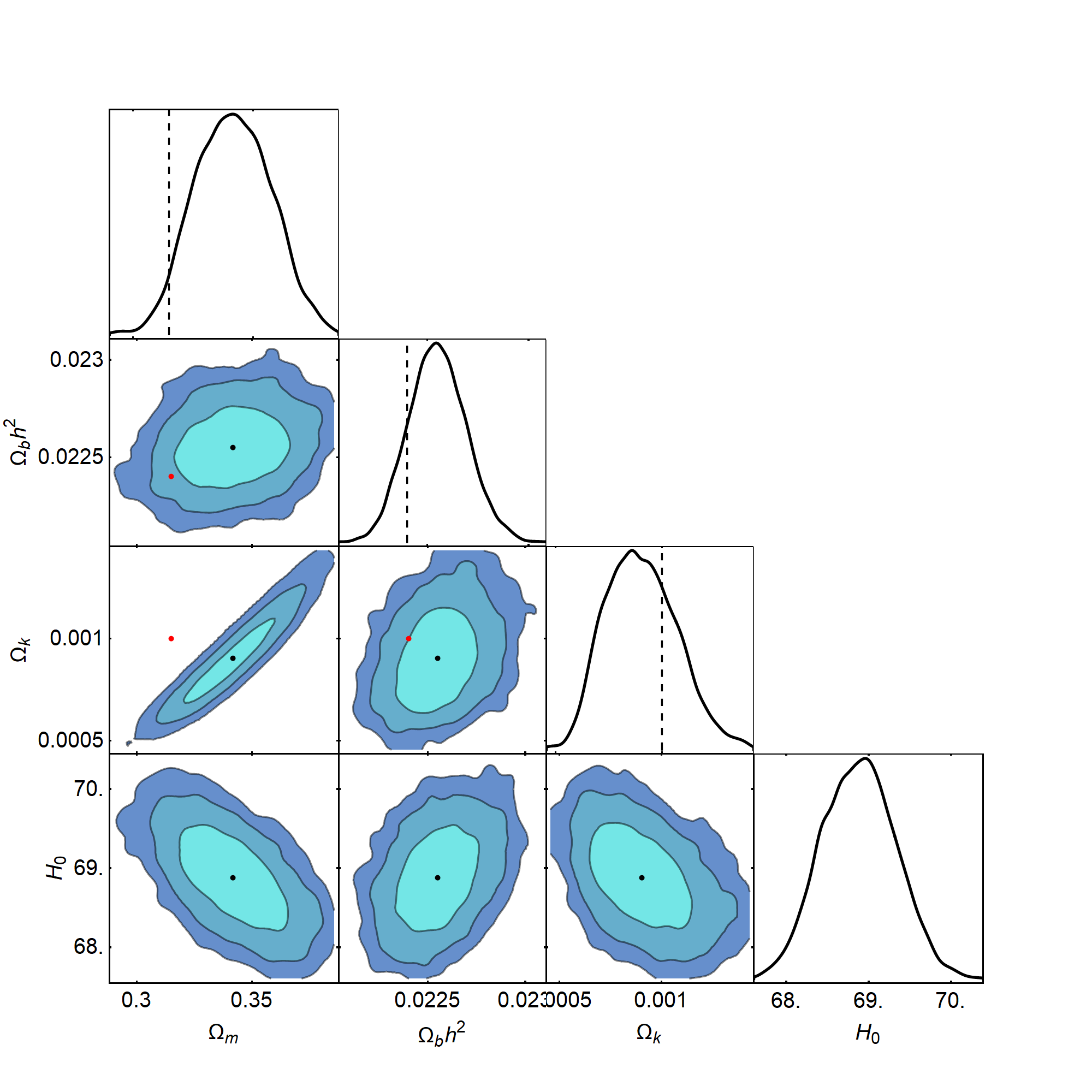}
    \includegraphics[width=0.497\textwidth]{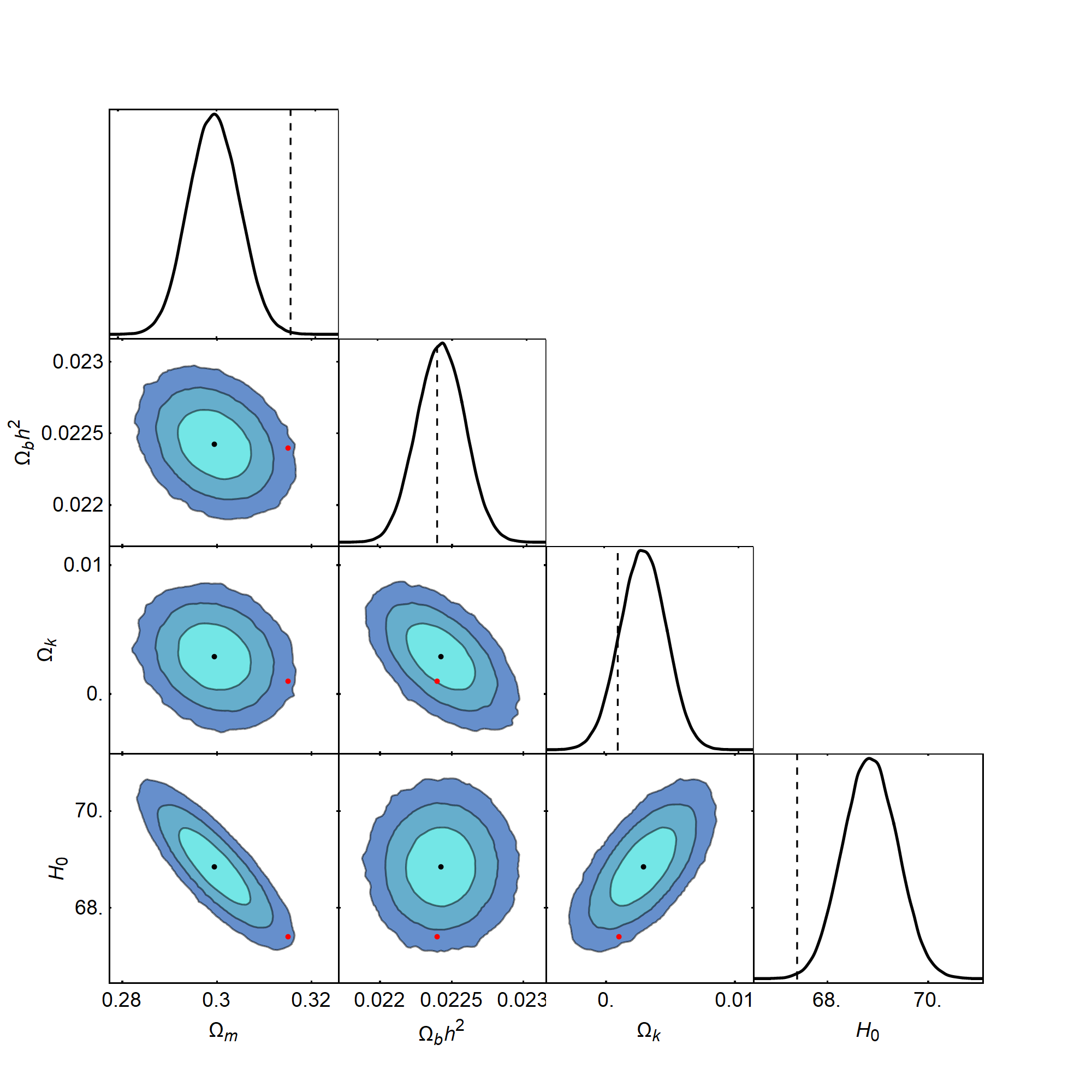}
    \caption{The $68.3\%$, $95.5\%$ and $99.7\%$ confidence contours for the GREAT (left panel) and \lcdm (right panel) models respectively, including all data and the Riess $H_0$ prior. The red points/dashed lines correspond to the Planck best-fit $(\Omega_{m,0},\Omega_{b,0}h^2,\Omega_{k,0},H_0)=(0.315,0.0224,0.001,67.4)$, where $H_0$ is given in units of $\mathrm{km}\,\mathrm{s}^{-1}\,\mathrm{Mpc}^{-1}$.}
    \label{fig:contours_Riess}
\end{figure*}

\begin{table*}[!t]
\begin{tabular}{cccccccc}
  \hline
  Model & $\Omega_{m,0}$ & $\Omega_{b,0}h^2$ & $\Omega_{k,0}$ & $H_0$ & $\chi^2_{min}$ & $\log Z(1)$ &$ \Delta \log Z(1)_{\Lambda,i}$\\ \hline
  $\Lambda$CDM & $0.2995\pm 0.0051$ & $0.0224\pm 0.0002$ & $0.0029\pm0.0017$ & $68.85\pm0.53$ & $1088.79 $& -557.588 & $0$ \\
  GREAT& $0.3350\pm 0.0155$ & $0.0225\pm 0.0001$ & $0.0008\pm 0.0002$ & $68.98\pm 0.44$ & $1083.39 $& -557.974 & $0.386$\\
  \hline
\end{tabular}
\caption{Here we present the results of the MCMC analysis when we include all the available data and the Riess $H_0$ prior, as discussed in the previous sections. In particular, we show the mean values, $1\sigma$ errors of the parameters for the GREAT and \lcdm models respectively, along with the minimum $\chi^2$ and the log-evidence $\log Z(1)$, see appendix~\ref{sec:app} and the difference of the log-evidence with respect to the \lcdm model $ \Delta \log Z(1)_{\Lambda,i}\equiv\log Z(1)_{\Lambda}-\log Z(1)_{i}$. The latter give a Bayes ratio of $B_{\Lambda, G}=\exp\left[\Delta \log Z(1)_{\Lambda,G}\right]=\exp\left(0.386\right)\sim 1.47$, thus resulting in the two models being considered statistically equivalent according to the Jeffreys' scale \cite{Nesseris:2012cq}. Note that $H_0$ is given in units of $\mathrm{km}\,\mathrm{s}^{-1}\,\mathrm{Mpc}^{-1}$.\label{tab:vals_Riess}  }\vspace{0.5cm}
\end{table*}

Similarly, in the figures we will always show the $68.3\%$, $95.5\%$ and $99.7\%$ confidence contours for the GREAT (left panel) and \lcdm (right panel) models respectively. In all cases, the black points will correspond to the mean values of the parameters from the MCMC, the blue shaded regions will be the  confidence levels, while the red points will correspond to the Planck 2018 best-fit $(\Omega_{m,0},\Omega_{b,0}h^2,\Omega_{k,0},H_0)=(0.315,0.0224,0.001,67.4)$, with $H_0$ given in units of $\mathrm{km}\,\mathrm{s}^{-1}\,\mathrm{Mpc}^{-1}$.

First, we consider the case when we include all of the data, except the priors on $H_0$, as they may be in some tension with other data \cite{Efstathiou:2020wxn,Efstathiou:2021ocp}. In particular, in Table~\ref{tab:vals_no_H0} we provide the results for the relevant parameters of the two models and as can be seen, in this case the thermodynamic MCMC analysis gives a Bayes ratio of $B_{\Lambda, G}=\exp\left[\Delta \log Z(1)_{\Lambda,G}\right]=\exp\left(-9.006\right)\sim 1/8150$, thus resulting in very strong evidence in favor of the GREAT model according to the Jeffreys' scale \cite{Nesseris:2012cq}. The corresponding confidence contours are given in Fig.~\ref{fig:contours_no_H0}.

\snt{As this case gives the strongest result in favor of GREAT, we also analyse in more detail what piece of experimental data is contributing to this improvement over the \lcdm model. In particular, as can be seen in Table \ref{tab:vals_no_H0_chi2s}, there is a difference of $\chi^2$ of $\sim4.3$ between \lcdm ($\chi^2=1075.63$) and GREAT ($\chi^2=1071.35$) and the different datasets contribute in different ways. Specifically, the main effect comes from the CMB data ($\delta \chi^2\sim 4.21$) and to a much lesser degree from the $H(z)$ data ($\delta \chi^2\sim 0.42$). On the other hand the BAO favor \lcdm slightly ($\delta \chi^2\sim -0.4$) and the $\chi^2$ for the SnIa is practically the same.}

Second, we also consider the case where we include all the data, along with the Riess $H_0$ prior of Ref.~\cite{Riess:2020fzl}. In Table~\ref{tab:vals_Riess} we provide the results for the relevant parameters of the two models and as can be seen, the thermodynamic integration gives a Bayes ratio of $B_{\Lambda, G}=\exp\left[\Delta \log Z(1)_{\Lambda,G}\right]=\exp\left(0.386\right)\sim 1.47$, thus resulting in the two models being considered statistically equivalent according to the Jeffreys' scale, see Table~\ref{tab:Jef} and Ref.~\cite{Nesseris:2012cq}. The corresponding confidence contours are given in Fig.~\ref{fig:contours_Riess}.

Furthermore, in Fig.~\ref{fig:contoursw0wa_Riess} we show the confidence contours for the $w_0$, $w_a$ parameters of the $w_0w_a$CDM model, which has an equation of state $w(a)=w_0 + w_a(1-a)$ \cite{Chevallier:2000qy, Linder:2002et}. As can be seen, as predicted by GREAT, the point $(w_0,w_a)=(-0.946,-0.318)$ \cite{Garcia-Bellido:2021idr}, denoted by an orage star in the plot, is very close to the best-fit of the model and in good agreement with observations in this case.

\begin{figure*}[!t]
    \centering
    \includegraphics[width=0.497\textwidth]{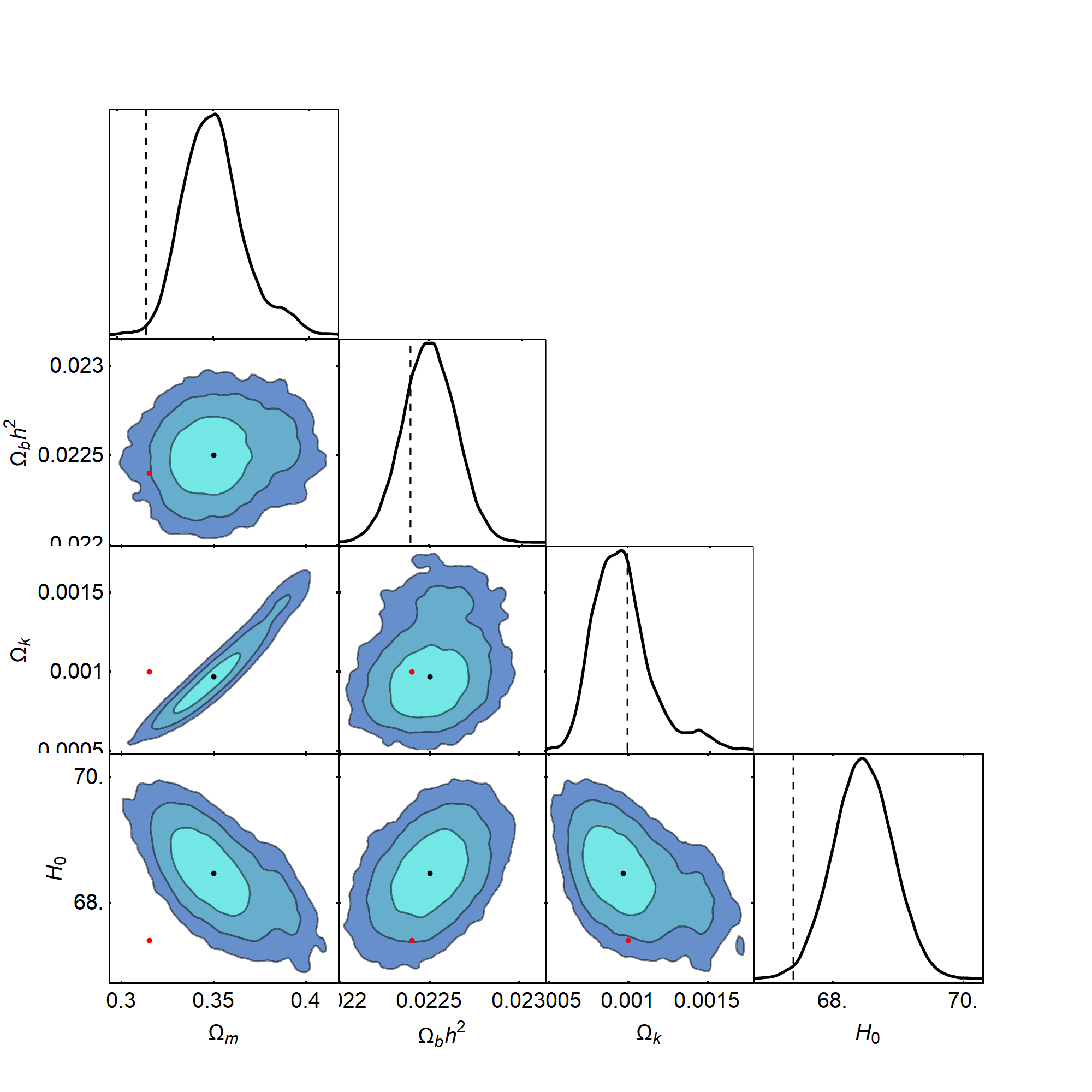}
    \includegraphics[width=0.497\textwidth]{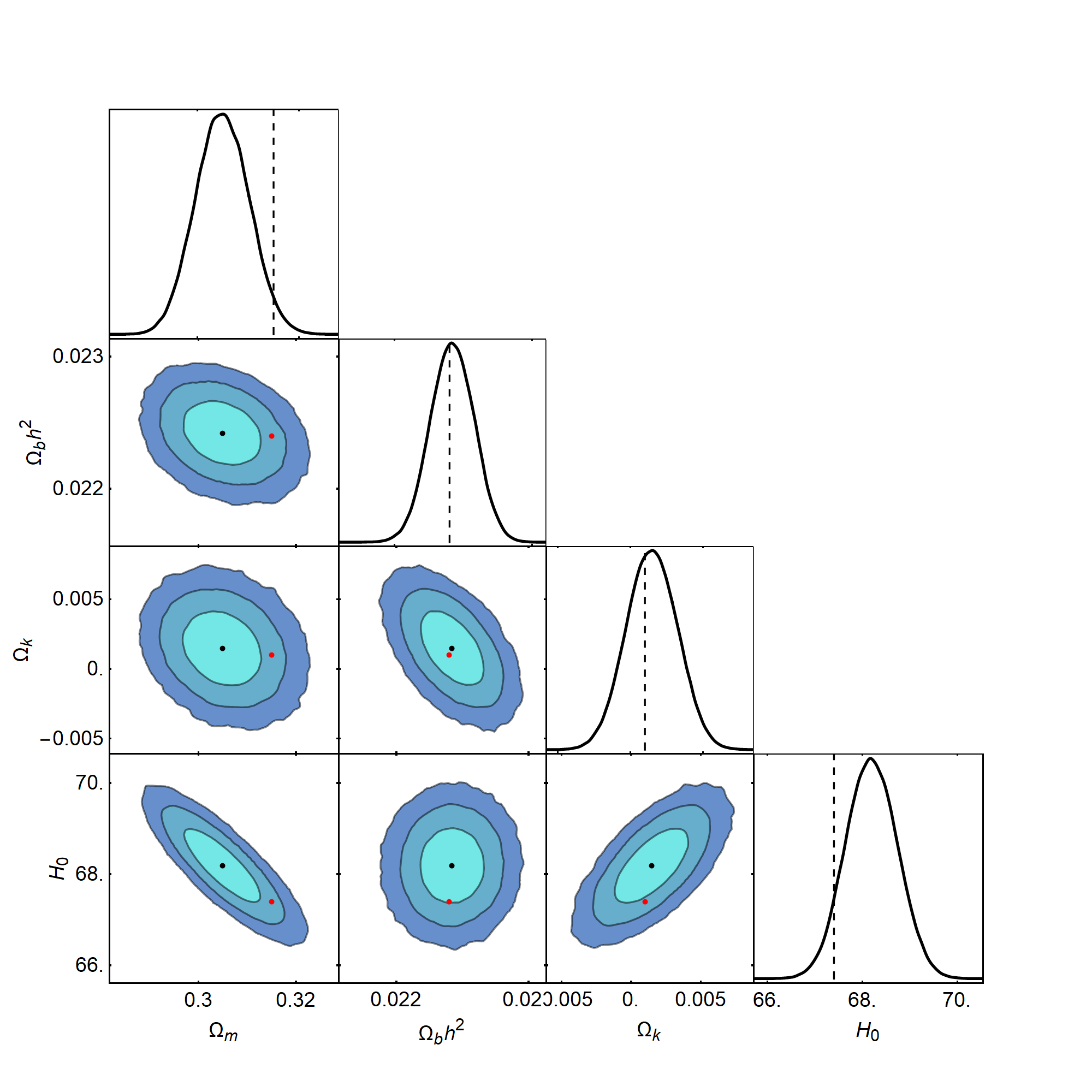}
    \caption{The $68.3\%$, $95.5\%$ and $99.7\%$ confidence contours for the GREAT (left panel) and \lcdm (right panel) models respectively, including all data and the TRGB prior on $H_0$. The red points/dashed lines correspond to the Planck best-fit $(\Omega_{m,0},\Omega_{b,0}h^2,\Omega_{k,0},H_0)=(0.315,0.0224,0.001,67.4)$, where $H_0$ is given in units of $\mathrm{km}\,\mathrm{s}^{-1}\,\mathrm{Mpc}^{-1}$.}
    \label{fig:contours_TRGB}
\end{figure*}

Finally, we also consider the case with all the data and the TRGB $H_0$ prior of Ref.~\cite{Freedman:2020dne}. In Table~\ref{tab:vals_TRGB} we provide the results for the relevant parameters of the two models and as can be seen, the thermodynamic integration gives a Bayes ratio of $B_{\Lambda, G}=\exp\left[\Delta \log Z(1)_{\Lambda,G}\right]=\exp\left(-0.373\right)\sim 0.689$, thus resulting in the two models being considered statistically equivalent according to the Jeffreys' scale, see Table~\ref{tab:Jef} and Ref.~\cite{Nesseris:2012cq}. The corresponding confidence contours are given in Fig.~\ref{fig:contours_TRGB}.

\section{Conclusions}
\label{sec:conclusions}

\begin{figure}[!t]
    \centering
    \includegraphics[width=0.45\textwidth]{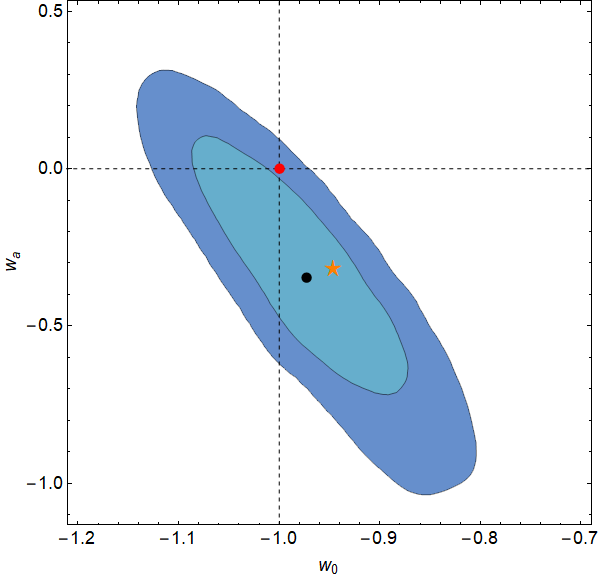}
    \caption{The $68.3\%$ and $95.5\%$ confidence contours for the CPL model for the $w_0$, $w_a$ parameters, when including all data and the Riess prior. The black dot corresponds to the best-fit value, the red dot to the \lcdm model and the orange star to the prediction of GREAT $(w_0,w_a)=(-0.946,-0.318)$ \cite{Garcia-Bellido:2021idr}.}
    \label{fig:contoursw0wa_Riess}
\end{figure}

\begin{table*}[!t]
\begin{tabular}{cccccccc}
  \hline
  Model & $\Omega_{m,0}$ & $\Omega_{b,0}h^2$ & $\Omega_{k,0}$ & $H_0$ & $\chi^2_{min}$ & $\log Z(1)$ &$ \Delta \log Z(1)_{\Lambda,i}$\\ \hline
  $\Lambda$CDM & $0.3047 \pm 0.0052 $ & $0.0224\pm  0.0001 $ & $0.0015\pm 0.0017 $ & $68.20\pm 0.54 $ & $1076.23 $ & -550.484 & $0$ \\
  GREAT & $0.3502 \pm 0.0157 $ & $0.0225\pm 0.0001 $ & $0.0010\pm 0.0002 $ & $68.46\pm 0.45 $ & $ 1071.74 $ & -550.111 & -0.373 \\
  \hline
\end{tabular}
\caption{Here we present the results of the MCMC analysis when we include all the available data and the TRGB $H_0$ prior, as discussed in the previous sections. In particular, we show the mean values, $1\sigma$ errors of the parameters for the GREAT and \lcdm models respectively, along with the minimum $\chi^2$ and the log-evidence $\log Z(1)$, see appendix~\ref{sec:app} and the difference of the log-evidence with respect to the \lcdm model $ \Delta \log Z(1)_{\Lambda,i}\equiv\log Z(1)_{\Lambda}-\log Z(1)_{i}$. The latter give a Bayes ratio of $B_{\Lambda, G}=\exp\left[\Delta \log Z(1)_{\Lambda,G}\right]=\exp\left(-0.373\right)\sim 0.689$, thus resulting in the two models being considered statistically equivalent according to the Jeffreys' scale \cite{Nesseris:2012cq}. Note that $H_0$ is given in units of $\mathrm{km}\,\mathrm{s}^{-1}\,\mathrm{Mpc}^{-1}$.\label{tab:vals_TRGB}  }
\end{table*}

The matter and energy content of the universe can only be inferred indirectly from the light that reaches us from distant sources which are affected by the expansion of the universe. It is therefore needed to interpret those measurements in the context of a given framework. We have assumed a spatially-curved, homogeneous and isotropic universe and determined the parameters of the model that best fit the currently available data, from CMB to Large Scale Structure (LSS), SnIa and local rate of expansion measurements.

The origin of the present acceleration of the universe is still a mystery. So far, the best model that fits the data is $\Lambda$CDM, where the acceleration is driven by a cosmological constant, whose origin is completely unknown, and whose value cannot be accounted for by quantum physics. In this paper we have explored the possibility, outlined in Ref.~\cite{Garcia-Bellido:2021idr}, that the present acceleration is driven by the growth of entropy associated with the cosmological horizon, a term in the action that inevitably appears in general relativity for fluids far from equilibrium~\cite{Espinosa-Portales:2021cac}. Such a surface term could give rise to the observed acceleration without the need to invoke any cosmological constant or extra fields. 

Whether this entropic force is all that is needed to explain the present cosmological observations was the main aim of this research. We are aware that there could be extra entropic contributions to the acceleration of the universe coming from bulk entropy growth processes, e.g. associated with the merging and mass accretion of black holes at the centers of galaxies, or the formation of the cosmic web itself, a highly ordered system very different from the uniform gas from which it arose.

We have performed a series of tests of the GREA theory with observations of the CMB, LSS and SnIa, and added to these the recent determinations of the present rate of expansion $H_0$, by Riess et al. (Cepheids) and Freedman et al. (TRGB). We find that, in the absence of an extra prior on $H_0$, the GREA theory fairs significantly better than $\Lambda$CDM, with log of the Bayes factor of order 9 in favor of GREA, a feat that has never been reached up to date for any alternative to $\Lambda$CDM. When including the Cepheids or the TRGB priors, the $\Delta\chi^2$ and Bayes evidence is uninformative, with $|{\rm log\,Bayes}| < 1$.

Moreover, when extending $\Lambda$CDM beyond a constant $\Lambda$ into $w_0w_a$CDM, we find that GREA theory predictions $(w_0,w_a)=(-0.946,-0.318)$ fall very near the best fit values, see Fig.~\ref{fig:contoursw0wa_Riess}, while $\Lambda$CDM is at the edge, within the 2-sigma contour. In the future, such contours will be significantly reduced and one will be able to differentiate easily between the two alternatives.

We conclude that GR entropic acceleration is a serious contender as a theory of the late universe and expect future measurements by CMB-S4, Euclid, DESI and LSST, to provide a definite conclusion. The realization that there is no need for a cosmological constant and that known physics (General Relativity, Quantum Mechanics and Thermodynamics) is all that is needed to explain the late time observations, could change our way we understand the origin and evolution of our Universe.

\textit{Numerical Analysis Files}: The \texttt{Mathematica} codes used by the authors in the analysis of the paper can be found at \href{https://github.com/snesseris/GREAT-project}{https://github.com/snesseris/GREAT-project} \\

\section*{Acknowledgements}
The authors acknowledge support from the Research Project  PGC2018-094773-B-C32 and the Centro de Excelencia Severo Ochoa Program SEV-2016-0597. S.~N. also acknowledges support from the Ram\'{o}n y Cajal program through Grant No. RYC-2014-15843. The work of L.E.P. is funded by a fellowship from ``La Caixa" Foundation (ID 100010434) with fellowship code LCF/BQ/IN18/11660041 and the European Union Horizon 2020 research and innovation programme under the Marie Sklodowska-Curie grant agreement No. 713673.\\

\begin{appendix} 
\section{Thermodynamic integration for the Bayesian evidence  \label{sec:app}}
In order to estimate the evidence integral, we can use thermodynamic MCMC integration \cite{Beltran:2005xd,thermoint}. To do so, we define the evidence as a function of the inverse temperature $\beta=1/T$ as follows:
\be 
Z(\beta)=\int d^n x\,\mathcal{L}(x)^\beta \, p(x),
\ee 
where $\mathbf{x}$ are the $n$ parameters of the model, the likelihood is $\mathcal{L}(x)$ and finally the prior $p(x)$ is assumed to be normalized, i.e. $\int d^n x\,p(x)=1$. Then, the actual Bayes factor, i.e. the evidence, of the model is just $Z(1)$. Furthermore, it is easy to show that
\ba
\frac{d \ln Z}{d\beta}&=& \frac{1}{Z(\beta)}\int d^nx\,(\ln\mathcal{L})\, \mathcal{L}(x)^\beta p(x)\nn \\
&=& \langle \ln\mathcal{L}\rangle_\beta, 
\ea 
where $\langle \ln\mathcal{L}\rangle_\beta$ is the average log-likelihood  over the posterior at an inverse temperature $\beta$. Since $Z(0)=1$, as the prior is normalized, then we get
\be 
\ln Z(1)=\int_0^1 d\beta\,\langle \ln\mathcal{L}\rangle_\beta.\label{eq:Z1}
\ee 

The integral in the last expression can be calculated  by estimating the average log-likelihood of each chain at a given inverse temperature and then performing the integral numerically. In practice we use an irregular grid with step size $\beta_i=\left(\frac{i}{N}\right)^5$, where $N$ is the number of steps in the grid. For the actual tempered MCMCs, we use the numerical code of one of the authors.

\end{appendix}

\bibliography{great}

\end{document}